\documentclass[aip,jcp,reprint,showkeys,superscriptaddress,a4paper,amsmath]{revtex4-1}
\usepackage{graphicx}
\usepackage{epsfig}
\usepackage{amsmath}
\usepackage{graphics}
\usepackage{latexsym}
\usepackage{amssymb}
\usepackage{inputenc}
\usepackage{color}

\newcommand{\la}{\left\langle}
\newcommand{\ra}{\right\rangle}

\newcommand{\angstrom}{\textup{\AA}}

\def \figwidth {\columnwidth}

\begin{document}

\title{Wetting transitions of polymer solutions: Effects of chain length and chain stiffness}

\author{Jiarul Midya}
\affiliation{Theoretical Soft Matter and Biophysics, Institute for Advanced Simulation and Institute of Complex Systems, Forschungszentrum J{\"u}lich, 52425 J{\"u}ulich, Germany}
\affiliation{Institute of Physics, Johannes Gutenberg University Mainz, Staudingerweg 7, 55128 Mainz, Germany}
\author{Sergei A. Egorov}
\affiliation{Department of Chemistry, University of Virginia, Charlottesville, Virginia 22901, USA}
\affiliation{Institute of Physics, Johannes Gutenberg University Mainz, Staudingerweg 7, 55128 Mainz, Germany}
\author{Kurt Binder}
\affiliation{Institute of Physics, Johannes Gutenberg University Mainz, Staudingerweg 7, 55128 Mainz, Germany}
\author{Arash Nikoubashman}
\affiliation{Institute of Physics, Johannes Gutenberg University Mainz, Staudingerweg 7, 55128 Mainz, Germany}
\email{anikouba@uni-mainz.de}
\begin{abstract}
Wetting and drying phenomena are studied for flexible and semiflexible polymer solutions via coarse-grained molecular dynamics simulations and density functional theory calculations. The study is based on the use of Young's equation for the contact angle, determining all relevant surface tensions from the anisotropy of the pressure tensor. The solvent quality (or effective temperature, equivalently) is varied systematically, while all other interactions remain unaltered. For flexible polymers, the wetting transition temperature $T_{\rm w}$ increases monotonically with chain length $N$, while the contact angle at temperatures far below $T_{\rm w}$ is independent of $N$. For semiflexible polymer solutions, $T_{\rm w}$ varies non-monotonically with the persistence length: Initially, $T_{\rm w}$ increases with increasing chain stiffness and reaches a maximum, but then a sudden drop of $T_{\rm w}$ is observed, which is associated with the isotropic-nematic transition of the system. 
\end{abstract}
\maketitle

\section{Introduction}
Wetting phenomena in polymer solutions are important from a basic scientific point of view as well as in the context of many applications in the area of materials science, e.g., coating of surfaces to modify their adhesive properties, protecting surfaces against corrosion, modifying lubrication properties of surfaces, etc. It has been known for more than 200 years that wetting of solid surfaces is controlled by the competition of the vapor-liquid (vl) surface tension, $\gamma_{\rm vl}$, and the difference of the wall-vapor (wv) surface tension, $\gamma_{\rm wv}$, and the wall-liquid (wl) surface tension, $\gamma_{\rm wl}$. When $\gamma_{\rm vl} < \gamma_{\rm wv} - \gamma_{\rm wl}$, the surface is coated by a mesoscopic liquid layer, while otherwise Young's equation \cite{TYoungPTRS1805} describes the contact angle $\theta$ of droplets at equilibrium
\begin{align}
    \cos(\theta) = (\gamma_{\rm wv} - \gamma_{\rm wl})/\gamma_{\rm vl} .
    \label{Eq:Young}
\end{align}
These droplets attached to the surface ideally have the shape of a capped sphere.

The above description can be readily carried over to two-component fluid systems, where analogous phenomena can occur when phase separation is found between two phases, poor or rich in one of the components, which play the role of the vapor or the liquid in Eq.~(\ref{Eq:Young}).\cite{PGdeGennesRevModPhys1985, S.Dietrich:book1988, EMBlokhuisCOCIS1996,DBonnRepProgPhys2001, DBonnRevModPhys2009} In this way, Eq.~(\ref{Eq:Young}) is used for binary (A, B) polymer blends, that separate into A-rich and B-rich phases,\cite{ISchmidtJPhysFrance1985, JKleinScience1990,USteinerScience1992, RALJonesPolymer1994, MGeogheganProgPolymSci2003} and for polymer solutions which may separate into solvent-rich and polymer-rich coexisting phases.\cite{HNakanishiJChemPhys1983, JForsmanJChemPhys2002, MMuellerJChemPhys2003, PBrykJChemPhys2004} It is of particular interest that the chain length $N$ enters as a control parameter which can be varied, keeping all interactions on the scale of monomeric units unchanged. \cite{KBinderAdvPolymSci1994} In symmetric polymer blends ($N_{\rm A} = N_{\rm B} = N$), one expects from mean-field theory that the interface tension between the coexisting phases of the unmixed blend $\gamma_{\rm vl}$ is of order $k_{\rm B}T/N^{1/2}$,\cite{PJFlory:book1953, PGdeGennes:book1979} while $\gamma_{\rm wv}$ and $\gamma_{\rm wl}$ can still be of the order of the thermal energy $k_{\rm B}T$, as for small molecules systems. For polymer solutions, where unmixing occurs at temperatures well below the theta temperature,\cite{PJFlory:book1953, PGdeGennes:book1979} $\gamma_{\rm vl}$ is also small close to the critical point (of order $k_{\rm B}T/N^{1/4}$ or even smaller).\cite{BWidomJStatPhys1988, SEndersJChemPhys1995} Due to the existence of such a control parameter, polymer solutions are interesting model systems for investigating fundamental aspects of wetting phenomena.

Particular interest in wetting phenomena was ignited by the discovery \cite{JWCahnJChemPhys1977} that varying the temperature can induce a phase transition , i.e., a singularity of the surface excess free energy of the fluid due to the wall, from a state of partial wetting [$\cos(\theta) < 1$] to complete wetting [$\cos(\theta)=1$] at a transition temperature $T_{\rm w}$. While wetting transitions for small molecule systems have been studied extensively by experiments, analytical theories and computer simulations,\cite{S.Dietrich:book1988, DBonnRepProgPhys2001, KBinderAnnRevsMaterRes2008} work on polymer solutions is still rather scarce: Experiments are hampered by sample preparation problems such as polydispersity of the chain lengths,\cite{MBuzzacchiPhysRevLett2006} etc.; analytical theories and simulations are difficult due to the multitude of length scales describing already the conformation of a single large macromolecule, ranging from the monomeric units ($\sim \angstrom$) to the scale of the persistence length ($\sim {\rm nm}$) and finally the end-to-end distance of the macromolecule ($\sim 10-100\,{\rm nm}$).\cite{KBinder:book1995}

In an interesting early attempt,\cite{JKleinMacromolecules1982} Klein and Pincus extended the Cahn description\cite{JWCahnJChemPhys1977} of the effects of a wall on a fluid exhibiting vapor-liquid like phase separation to a solution of flexible polymers in the poor solvent regime. To include the effects of the wall (located at $z=0$), the free energy functional was supplemented by a ``bare" surface term, $f_{\rm s, {\rm bare}}(\phi_{\rm s})$, which depends on the local polymer concentration $\phi_{\rm s}$ at the surface only. The bulk free energy density was taken from Flory-Huggins theory.\cite{KBinderAdvPolymSci1994, PJFlory:book1953, PGdeGennes:book1979} Assuming an adsorbed film with thickness $d$ at least of the order of the chain radius, and conditions where $\phi_{\rm s}$ exceeds the concentration at the liquid-like branch of the coexistence curve, $\phi_{\rm l}$, it is found that the concentration profile $\phi(z)$ decays from $\phi_{\rm s}$ at the surface to a very small value $\phi'$ far from the wall in two steps: First, $\phi(z)$ exhibits a fast power law decay toward $\phi_{\rm l}$, followed by a flat region for $z < d$, and then an exponential decay toward $\phi'$. However, the approach of $\phi$ toward the vapor-like branch of the coexistence curve $\phi_{\rm v}$ was not discussed in Ref.~\citenum{JKleinMacromolecules1982} yet (one would expect a logarithmic divergence of $d$, ``complete wetting'' \cite{S.Dietrich:book1988}, if the temperature is in between the wetting transition temperature and the critical temperature of the solution \cite{JWCahnJChemPhys1977}). A closely related treatment was then formulated for binary polymer blends, \cite{ISchmidtJPhysFrance1985, RALJonesPolymer1994, HNakanishiJChemPhys1983} where complete wetting was analyzed as well as the possibility of critical and tricritical wetting. Wetting in polymer solutions was then considered again much later using the Scheutjens-Fleer lattice version of self-consistent field theory\cite{JMHMScheutjensMacromolecules1985} by Leermakers et al.\cite{FAMLeermakersLangmuir1997}, while the most recent and comprehensive treatment along the lines of Refs.~\citenum{ISchmidtJPhysFrance1985, JWCahnJChemPhys1977, JKleinMacromolecules1982} was given by Dolinnyi.\cite{AIDolinnyiColloidJ2020, AIDolinnyiColloidJ2021_1, AIDolinnyiColloidJ2021_2} The regime of very long chains ($N \to \infty$) close to the unmixing critical point of the bulk has been considered most extensively to date; there, all bulk properties (including the interface tension $\gamma_{\rm vl}$, which in this approach coincides with Widom's\cite{BWidomJStatPhys1988} description) are described by simple power laws in terms of the distance from the critical point; the corresponding prefactors are also simple powers of $N$. In that region, wetting phase diagrams are presented, and the conditions for the occurrence of critical and tricritical wetting are derived, similar to the earlier studies of polymer blends. \cite{ISchmidtJPhysFrance1985, RALJonesPolymer1994, HNakanishiJChemPhys1983} However, the regime close to criticality is difficult to study already for bulk polymer solutions;\cite{SEndersJChemPhys1995, AAPovodyrevPhysicaA1999, MAAnisimovJChemPhys2005,MAAnisimovMolPhys2005} further, for not very long chains there is a wide range of temperatures where Ising-like critical behavior is observed rather than mean-field behavior.\cite{JMidyaJChemPhys2019} This Ising-like behavior is not captured by any of the theories mentioned above.

An even more important caveat concerns the assumption of short range forces between the wall and the fluid particles, which is implicit in the description of surface effects via $f_{\rm s, bare}(\phi_{\rm s})$,\cite{S.Dietrich:book1988} as such short-ranged forces dominate only in rare cases.\cite{DBonnRepProgPhys2001} More common are surface potentials whose attractive part decays according to an inverse power law with the distance $z$ from the wall, namely $\propto z^{-3}$ if we deal with a surface of a bulk three dimensional substrate,\cite{S.Dietrich:book1988} or $\propto z^{-4}$ if we deal with a quasi-two-dimensional substrate. The latter case is realized by adsorption on, e.g., thin graphite films, stiff membranes, or surfaces coated by surfactant layers that control wettability.\cite{XXuACSNano2014} Note that for the case of short range forces between the fluid particles but long range wall-fluid forces, one expects quite generally that only first order wetting transitions are possible.\cite{REvansPANS2019} Hence, there is no need to discuss critical or tricritical wetting\cite{HNakanishiPhysRevLett1982} in such cases any further.

In the present work, we study the wetting transition for polymers in solution by molecular dynamics (MD) simulations\cite{allen:book:2017} using a coarse-grained model for which the bulk phase diagram has been established in earlier work.\cite{JMidyaJChemPhys2019} We systematically investigate the effect of chain length and chain stiffness, finding that the wetting transition temperature $T_{\rm w}$ increases monotonically with increasing chain length $N$, while the drying transition temperature is independent of $N$. In contrast, variation of the chain stiffness leads to a non-monotonic change of the wetting transition temperature, with an initial increase of $T_{\rm w}$ with increasing persistence length, followed by a sudden drop of $T_{\rm w}$ at the isotropic-nematic transition of the system. A related, but for technical reasons somewhat different coarse-grained model for flexible polymers is studied by density functional theory (DFT) in the Appendix.

\section{Model and Methods}
\label{sec:model}
We use a coarse-grained polymer model, where each chain is described as a sequence of $N$ spherical beads of diameter $\sigma$ and mass $m$. The solvent is modeled implicitly, and the solvent quality is varied through the effective monomer-monomer pair interaction
\begin{equation}
    U_{\rm mm}(r, \lambda) = \begin{cases}
    U_{\rm LJ}(r) + (1-\lambda)\varepsilon & , r \leq 2^{1/6}\sigma \\
    \lambda U_{\rm LJ}(r) & , \text{else} ,
    \end{cases}
\label{eq:Umm}
\end{equation}
with standard Lennard-Jones (LJ) potential
\begin{equation}
	U_{\rm LJ}(r) = \begin{cases}
	4\varepsilon\left[(\sigma/r)^{12} - (\sigma/r)^6\right] & , r \leq r_{\rm c} \\
	0 & , \text{else} .
	\end{cases}
	\label{eq:ULJ}
\end{equation}
In Eqs.~(\ref{eq:Umm}) and (\ref{eq:ULJ}) above, $r$ is the distance between two particles, $\varepsilon$ is the interaction strength, and $r_{\rm c} = 4\,\sigma$ is the cutoff radius of the interaction. Further, $U_{\rm LJ}$ is multiplied by a smoothing polynomial for $r \geq 3.5\,\sigma$, to gradually decrease both the force and potential to zero at the cutoff radius.\cite{midya:acsnano:2020} The dimensionless parameter $\lambda \geq 0$ controls the solvent quality, which worsens with increasing $\lambda$. Thus, the parameter $\lambda$ plays the role of an inverse effective temperature, $T_{\rm eff} \equiv 1/\lambda$, and good solvent conditions are recovered for $T_{\rm eff} \to \infty$. We always used $\lambda=0$ for the interactions between bonded monomers.

The monomers are bonded via the finitely extensible nonlinear elastic (FENE) potential
\begin{equation}
  U_{\rm bond}(r) = -\frac{1}{2} k r^2_0 \ln \left[1-\left(r/r_0\right)^2\right] \quad , r < r_0,
  \label{Eq:bondPoten}
\end{equation}
with maximum bond extension $r_0 = 1.5\,\sigma$, and spring constant $k=30\,\varepsilon/\sigma^2$.\cite{GSGrestPhysRevA1986} These values lead to an equilibrium bond length of $\ell_{\rm b} \approx 0.97\,\sigma$, which impedes unphysical bond crossing. 

Bending stiffness is incorporated by introducing a bond-bending potential
\begin{equation}
  U_{\rm bend}(\Theta_{ijk}) = \varepsilon_{\rm bend} [1-\cos(\Theta_{ijk})],
  \label{Eq:bendPoten}
\end{equation}
where $\varepsilon_{\rm bend}$ controls the rigidity of a chain, and $\Theta_{ijk}$ is the angle between two subsequent bond vectors, $\mathbf{r}_{ij}$ and $\mathbf{r}_{jk}$, connecting monomers $i$, $j$ and $k$ of a chain ($\Theta_{ijk} = 0$ when the three monomers lie on a line). The persistence length of the polymers is defined as $\ell_{\rm p} = -\ell_{\rm b}/\ln\la\cos(\Theta_{ijk})\ra$. For $\kappa \equiv \varepsilon_{\rm bend}/(k_{\rm B}T) \gtrsim 2$, the expression for $\ell_{\rm p}$ can be approximated by $\ell_{\rm p}/\ell_{\rm b} \approx \kappa$, where $k_{\rm B}$ is Boltzmann's constant.

If walls are present, they are modeled as smooth planar surfaces, with normal vectors in the $z$ direction, that interact with the monomers via a (10-4) Mie potential 
\begin{equation}
    U_{\rm wall}(z)=\begin{cases}
    \varepsilon_{\rm Mie}\left[\left(\sigma/z\right)^{10}- \left(\sigma/z\right)^{4}\right] & , z \leq z_{\rm c, Mie}\\
    0 & , \text{else} ,
   \end{cases}
   \label{Eq:Mie-potential}
\end{equation}
with cutoff distance $z_{\rm c, Mie}=5\,\sigma$. The strength of the potential is fixed at $\varepsilon_{\rm Mie}/(k_{\rm B}T) = 3$. We used this wall potential, as it was used by some of us in a related study on the adsorption of (semi)flexible polymers from solution under good solvent conditions,\cite{AMilchevSoftMatter2017} to which the present model corresponds for large $T_{\rm eff}$.

All MD simulations are performed in the $\mathcal{N}VT$ ensemble, with total number of particles $\mathcal{N}$ in an elongated box with linear dimensions $L_x=L_y=64\,\sigma$ and $L_z=128\sigma$ and volume $V=L_x \times L_y \times L_z$. The solvent quality is varied via $T_{\rm eff}$, while the thermodynamic temperature of the system is kept constant at $T=1.0\varepsilon/k_{\rm B}$ using a Langevin thermostat. Thus, only the strength of the attractive monomer-monomer contribution in Eq.~(\ref{eq:Umm}) is adjusted, while leaving all other interaction strengths (and thus the resulting bond length $\ell_{\rm b}$ and persistence length $\ell_{\rm p}$) fixed. The equations of motion are integrated using a velocity Verlet scheme with time step $\Delta t= 0.005 \tau$, $\tau = \sqrt{m \sigma^2/(k_{\rm B} T)}$ being the unit of time.\cite{Frenkel_Smit_Book} We used the HOOMD-blue software package (v. 2.9.3) to perform our simulations\cite{anderson:cms:2020} and rendered the snapshots with Visual Molecular Dynamics (VMD).\cite{WHumphreyJMolGraphics1996}

The interfacial tensions $\gamma$ are computed for each choice of $\lambda$, $N$, and $\kappa$ using the Kirkwood-Buff relation\cite{kirkwood:jcp:1949}
\begin{equation}
    \gamma = \frac{L_z}{2} \left(P_{zz} - \frac{P_{xx}+P_{yy}}{2}\right) ,
    \label{Eq:KirkwoodBuff}
\end{equation}
where the factor $1/2$ comes if there are two interfaces present in the system. The components of the pressure tensor, $P_{\alpha\beta}$, are calculated via the standard Clausius virial equation. 

The vapor-liquid surface tension of the polymer, $\gamma_{\rm vl}$, is determined by initially placing the chains in a slab at the center of the simulation box, with its two surface normals lying parallel to the $z$-axis. Periodic boundary conditions are applied in all directions, and the systems are equilibrated until coexistence of a low density vapor phase with a high density liquid phase has developed. The thickness of the slab is chosen large enough that the two vapor-liquid interfaces can be assumed as non-interacting with each other.

\begin{figure}[htbp!]
    \centering
    \includegraphics[width=\figwidth]{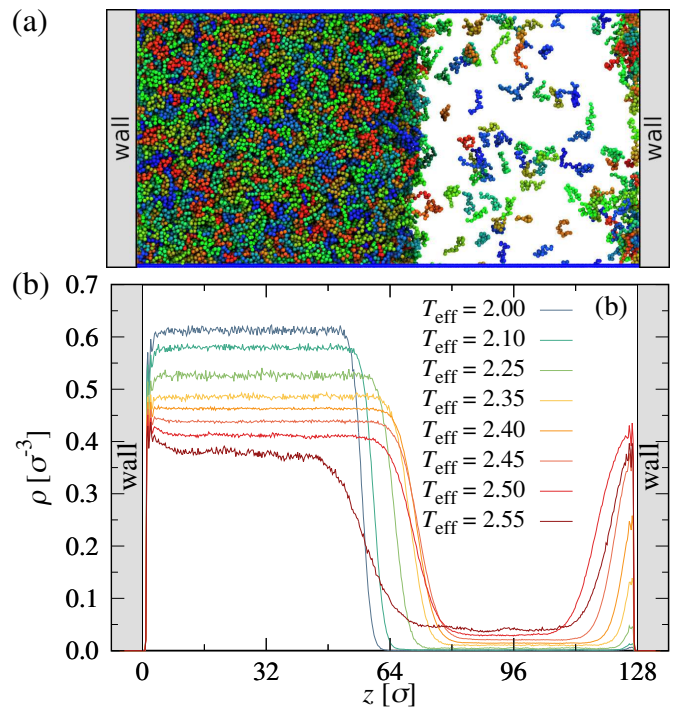}
    \caption{(a) Snapshot of flexible polymers ($\kappa = 0$) with length $N=16$ confined between two attractive walls at $T_{\rm eff}=2.40$ and with overall monomer density $\rho_{\rm tot} \simeq \rho_{\rm c} = 0.197\,\sigma^{-3}$, $\rho_{\rm c}$ being the critical density. (b) Corresponding monomer density distributions as functions of $z$ at various $T_{\rm eff}$, as indicated.}
    \label{Fig:snap+densProf_WVL}
\end{figure}

To determine $\gamma_{\rm wv}$ and $\gamma_{\rm wl}$, we perform additional simulations, where the polymer is confined between two attractive walls placed at $0$ and $L_z$, while periodic boundary conditions are applied along the $x$- and $y$-directions. Initial configurations are generated by placing a liquid slab of polymers next to one of the walls, while the other wall is occupied by a low density vapor phase. In this case, there are three well separated non-interacting interfaces (as $L_z$ is large enough), i.e., a wall-liquid, a liquid-vapor, and a vapor-wall interface [see Fig. \ref{Fig:snap+densProf_WVL}(a)]. Because the system is ``self-regulating'', the liquid and vapor phases have the densities precisely according to the coexistence curve in the bulk. Due to the adsorption at the wall-vapor interface, it is important to stay with the effective temperature $T_{\rm eff}$ in the regime of partial wetting. The amount of the liquid phase gets somewhat reduced as $T_{\rm eff}$ is raised, but the three interfaces are still far enough apart from each other to be treated as non-interacting [see Fig. \ref{Fig:snap+densProf_WVL}(b)]. Thus, the total interface tension of the entire system is
\begin{equation}
    \gamma_{\rm tot} = \gamma_{\rm wl} + \gamma_{\rm vl} + \gamma_{\rm wv} .
    \label{Eq:Split_gamma_tot}
\end{equation}

\begin{figure}[htbp!]
    \centering
    \includegraphics[width=\figwidth]{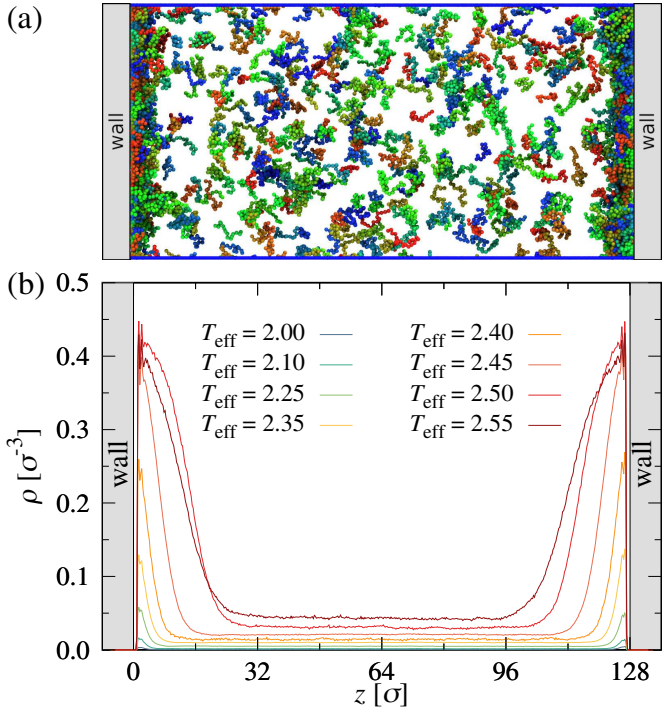}
    \caption{Same as Fig.~\ref{Fig:snap+densProf_WVL}, but with overall monomer density tuned such that the average density in the middle of the simulation box coincides with the vapor density at bulk vapor-liquid coexistence.}
    \label{Fig:snap+densProf+WV}
\end{figure}

To get $\gamma_{\rm wl}$ and $\gamma_{\rm wv}$ separately, a third simulation setup is needed, where the simulated state point is chosen such that the density in the bulk region coincides precisely with the coexistence density of the vapor branch (Fig.~\ref{Fig:snap+densProf+WV}). The choice of the total density $\rho_{\rm tot}$ in the simulation box then is a nontrivial issue due to adsorption at the walls: If $\rho_{\rm tot}$ is too small, then the vapor phase will be undersaturated, whereas a too large value of $\rho_{\rm tot}$ will result in a metastable supersaturated vapor. Both cases need to be avoided by iteratively tuning $\rho_{\rm tot}$ until the desired density in the vapor region is reached. From this setup, $\gamma_{\rm wv}$ can be calculated directly from Eq.~(\ref{Eq:KirkwoodBuff}), and $\gamma_{\rm wl}$ from Eq.~(\ref{Eq:Split_gamma_tot}). Hence, we have all the necessary input for Young's equation [Eq.~(\ref{Eq:Young})], allowing us to determine the contact angle and the wetting transition. Unlike the alternative method where large wall-attached droplets are studied, the present method is not affected by effects due to the line tension of the contact line where the vapor-liquid interface meets the wall.\cite{das:jpcm:2018}

\section{Results}
\label{sec:results}
\subsection{Bulk behavior}
\begin{figure}[htbp!]
    \centering
    \includegraphics[width=\figwidth]{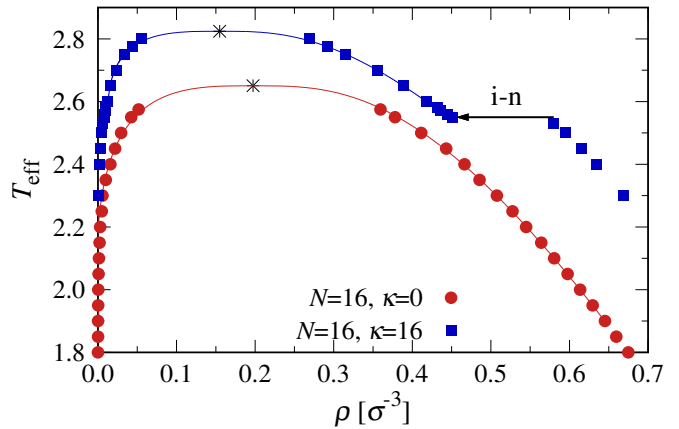}
    \caption{Phase diagrams of flexible ($\kappa = 0$) and semiflexible ($\kappa = 16$) polymers with chain length $N=16$ in the $\rho-T_{\rm eff}$ plane. Symbols indicate coexistence densities from our simulations, while the solid curves are fits to Eq.~(\ref{Eq:orderParamFit}). Critical points ($T_{\rm c}$, $\rho_{\rm c}$) are indicated by stars. For $\kappa=16$, isotropic-nematic (i-n) coexistence is indicated by an arrow.}
    \label{Fig:phase_diagram}
\end{figure}

Before discussing the wetting behavior of the polymers, we first give a brief overview on their bulk phase behavior (a more detailed discussion can be found in our previous study of this model\cite{JMidyaJChemPhys2019}). Figure~\ref{Fig:phase_diagram} shows the phase diagram of flexible ($\kappa=0$) and semiflexible ($\kappa=16$) polymers with $N=16$ in the $\rho-T_{\rm eff}$ plane. The left and right branches of the phase diagram represent the coexistence vapor ($\rho_{\rm v}$) and liquid ($\rho_{\rm l}$) densities, respectively. The difference in densities between the two phases, $\Delta \rho = \rho_{\rm l}-\rho_{\rm v}$, is the order parameter for the vapor-liquid transition, which vanishes at the critical point. The value of the critical temperature $T_{\rm c}$ and of the critical density $\rho_{\rm c}$ are estimated by using the universal scaling relations 
\begin{equation}
    \Delta \rho = A(T_{\rm c}-T_{\rm eff})^{\beta},
    \label{Eq:orderParamFit}
\end{equation}
and
\begin{equation}
    \rho_{\rm d} = \frac{\rho_{\rm l}+\rho_{\rm v}}{2}=\rho_{\rm c}+B(T_{\rm c}-T_{\rm eff}),
    \label{Eq:rectilinearDiaFit}
\end{equation}
where the constants $A$ and $B$ are material specific scaling parameters. We chose a critical exponent of $\beta=0.325$ for fitting our simulation data, assuming that our systems belong to the $3d$-Ising universality class due to the short range monomer-monomer interactions (see Sec.~\ref{sec:model}). To avoid confusion, we stress that we use here the language appropriate for a vapor-liquid transition throughout; but our model can be interpreted as a model for a polymer solution, if the solvent is implicitly accounted for by the effective interactions between monomeric units [Eqs.~(\ref{eq:Umm})].

For flexible polymers ($\kappa=0$), $T_{\rm c}$ increases monotonically with increasing $N$, whereas $\rho_{\rm c}$ decreases at the same time.\cite{KSilmoreMolPhys2017, JMidyaJChemPhys2019} This behavior is qualitatively in line with Flory-Huggins theory, which predicts that, in the limit of $N \to \infty$, $T_{\rm c}$ approaches the theta temperature as $T_{\rm c}-T_{\Theta} \propto N^{-1/2}$, while $\rho_{\rm c}$ decays to zero as $\rho_{\rm c} \propto N^{-1/2}$. This behavior also agrees qualitatively with the bulk DFT results shown in the Appendix. For semiflexible polymers ($\kappa > 0$), there is a monotonic increase in $T_{\rm c}$ and a decrease in $\rho_{\rm c}$ with increasing stiffness $\kappa$ at fixed chain length $N$.\cite{JMidyaJChemPhys2019} This behavior can be rationalized by considering that the mean-square radius of gyration of the polymers increases with increasing bending stiffness.

\begin{figure}[htbp!]
    \centering
    \includegraphics[width=\figwidth]{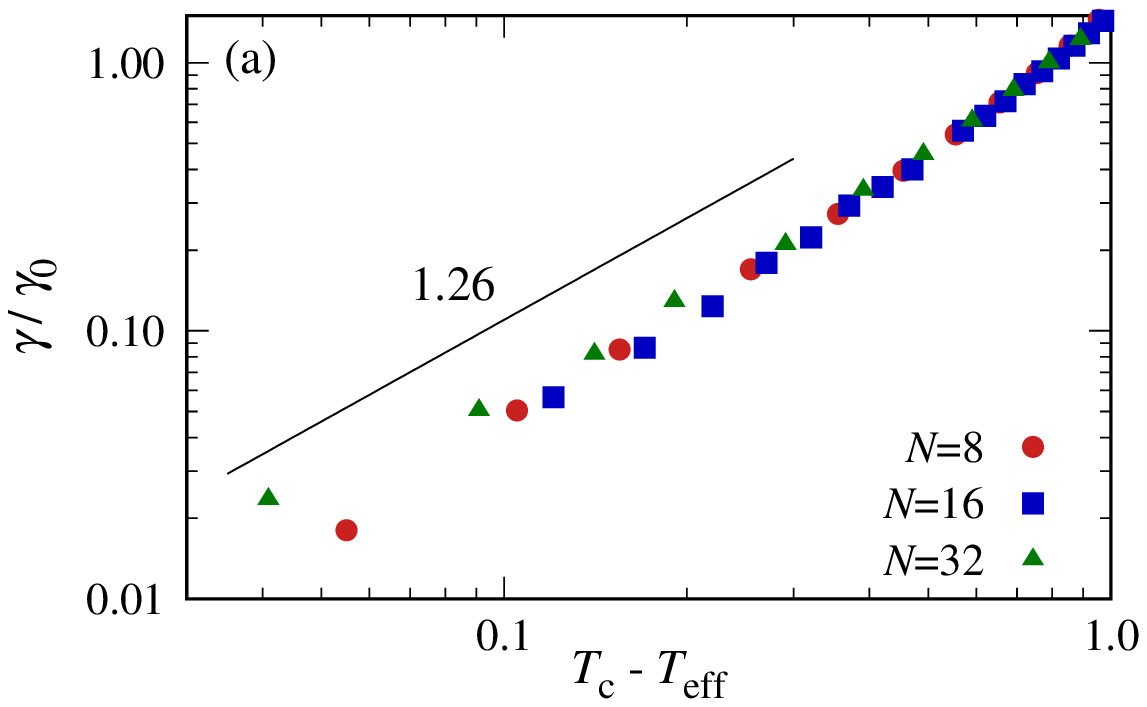}\\
    \includegraphics[width=\figwidth]{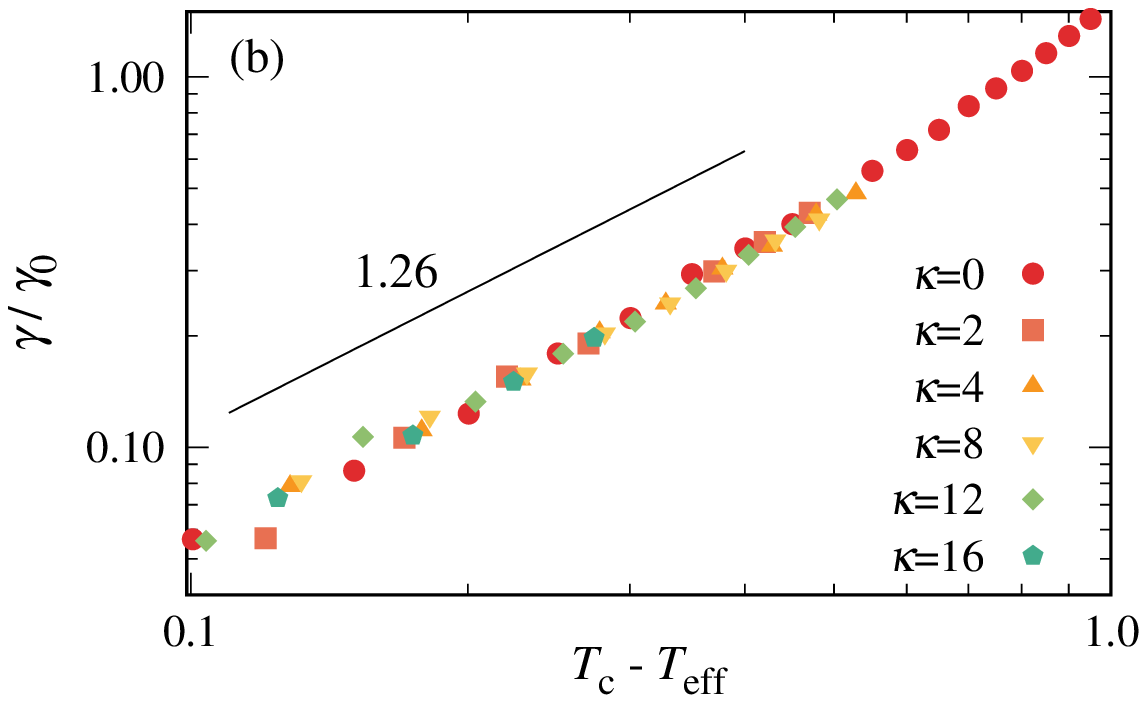}
    \caption{ (a) Normalized vapor-liquid surface tension $\gamma_{\rm vl}/\gamma_0$ vs $T_{\rm c}-T_{\rm eff}$ [Eq.~(\ref{eq:gammaVL})] for flexible polymers with three different lengths $N=8$, $16$, and $32$, in a log-log scale. The black line represents a power-law with exponent $\mu=1.26$. (b) Same as (a), but for different values of $\kappa=0$, $2$, $4$, $8$ and $16$ at fixed chain length $N=16$.}
    \label{Fig:norm_gamma_VL}
\end{figure}

In the vicinity of the critical point, the vapor-liquid surface tension follows the scaling relation
\begin{equation}
    \gamma_{\rm vl}=\gamma_0 (N) (T_{\rm c}-T_{\rm eff})^{\mu}
    \label{eq:gammaVL}
\end{equation}
with critical exponent $\mu=1.26$, according to the $3d$-Ising universality class. The critical amplitude $\gamma_0$ decreases with increasing chain length following an empirical power law for the investigated small range of $N$, $\gamma_0 \propto N^{-1/3}$. In Fig.~\ref{Fig:norm_gamma_VL}(a), we plot $\gamma_{\rm vl}/\gamma_0$ as functions of $T_{\rm c}-T_{\rm eff}$ for $N=8$, $16$ and $32$. The simulation data follow the expected scaling behavior in the vicinity of $T_{\rm c}$. Similar results for semiflexible polymers are presented in Fig.~\ref{Fig:norm_gamma_VL}(b) for different values of $\kappa=0$, $2$, $4$, $8$, $12$ and $16$ at fixed $N=16$. At fixed $N$, the value of $\gamma_0$ increases weekly with increasing $\kappa$. For different values of $\kappa$, $\gamma_{\rm vl}/\gamma_0$ vs $T_{\rm c}-T_{\rm eff}$ follows a master curve, which is consistent with the theoretically expected behavior, confirming that our system belong to the $3d$-Ising universality class as hypothesized.

\subsection{Wetting phenomena}
\begin{figure}[htbp!]
    \centering
    \includegraphics[width=\figwidth]{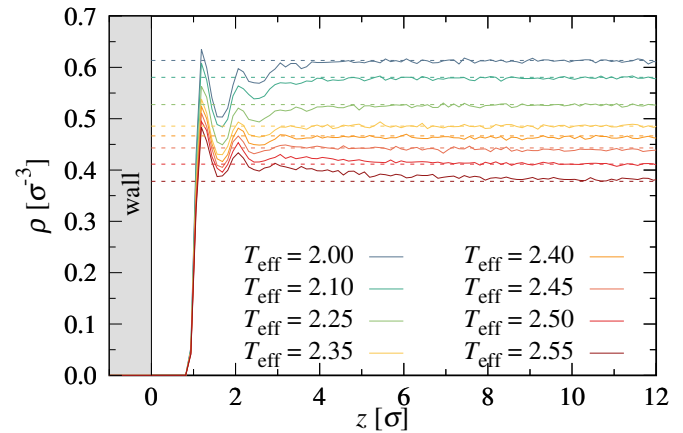}
    \caption{Density profiles of flexible polymers ($\kappa = 0$, $N=16$) near one of the attractive walls.}
    \label{Fig:zoomed_densProf+WVL}
\end{figure}

To study the wetting properties of the polymers, we confined them between two parallel attractive walls, located at $0$ and $L_z$ (see Sec.~\ref{sec:model}). Selected density distributions of monomers are presented in Figs. \ref{Fig:snap+densProf_WVL}(b) and \ref{Fig:zoomed_densProf+WVL} as function of distance $z$ (the overall monomer number density is fixed to $\rho_{\rm tot} \simeq \rho_{\rm c}$). The densities of the liquid and the vapor phase regions far away from the walls are identical to the corresponding coexistence densities in the bulk. Near the walls, however, the densities of both phases differ from their corresponding bulk coexistence densities. To better see the adsorption of polymers in the liquid region near the walls, we have plotted a closeup view of the density profiles in Fig.~\ref{Fig:zoomed_densProf+WVL}. 

\begin{figure}[htbp]
    \centering
    \includegraphics[width=\figwidth]{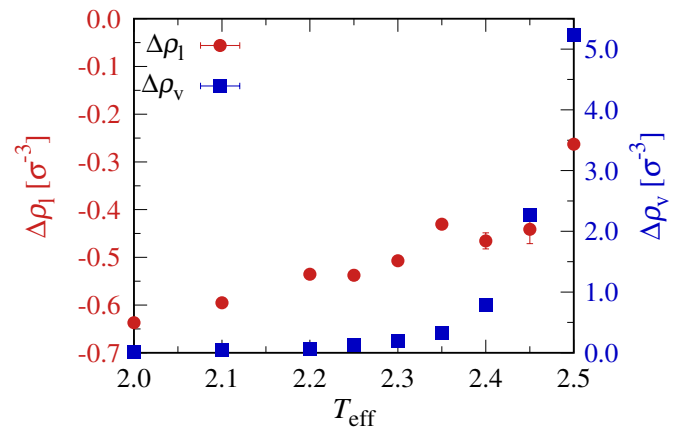}
    \caption{Excess density of the liquid ($\Delta \rho_{\rm l}$, left axis) and of the vapor phase ($\Delta \rho_{\rm v}$, right axis) vs $T_{\rm eff}$ for flexible polymers ($\kappa = 0$, $N=16$).}
    \label{Fig:excess}
\end{figure}

From the density profiles we can compute the excess density, $\Delta \rho_{\rm l} = \frac {1}{\sigma}\int_{0}^{d/2} {\rm d}z \left[\rho(z)-\rho_{\rm l}\right]$, where $d$ is the slab thickness, and $\rho_{\rm l}$ is the liquid density at bulk vapor-liquid coexistence (an analogous expression can be defined for the excess density of the vapor, $\Delta \rho_{\rm v}$). We find $\Delta \rho_{\rm l} < 0$ throughout, whereas $\Delta \rho_{\rm v}$ is positive and increases distinctly to a large but finite value at $T_{\rm w}$, where the transition from partial to complete wetting occurs (Fig.~\ref{Fig:excess}).

\begin{figure}[htbp!]
    \centering
    \includegraphics[width=\figwidth]{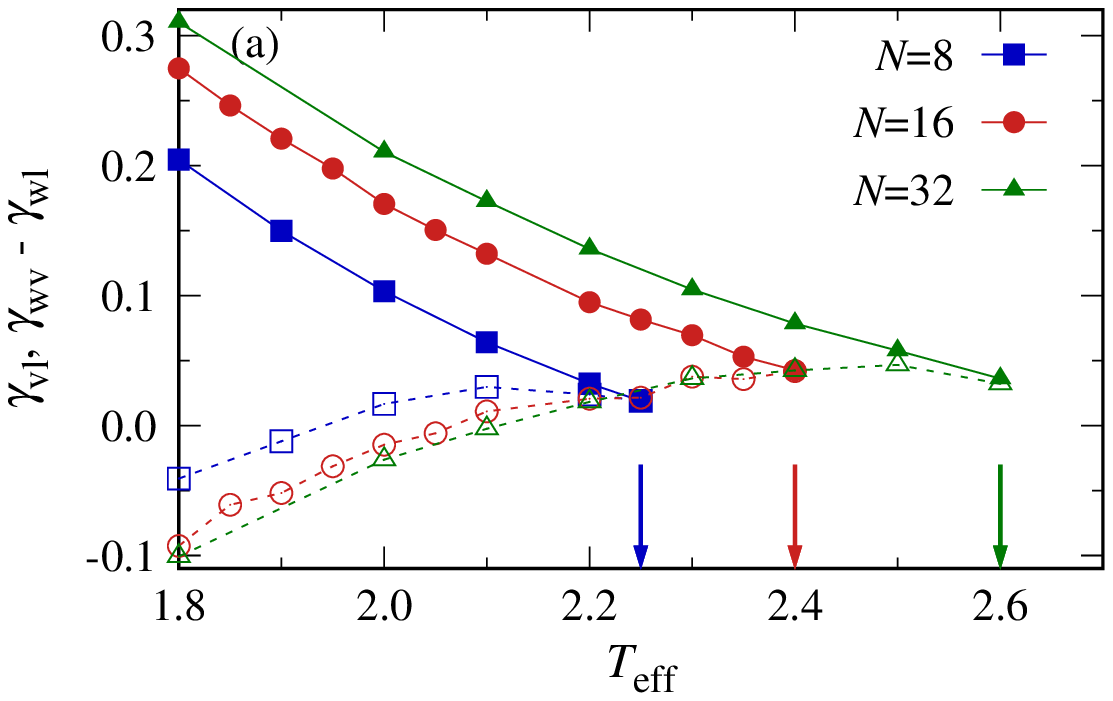}\\
    \includegraphics[width=\figwidth]{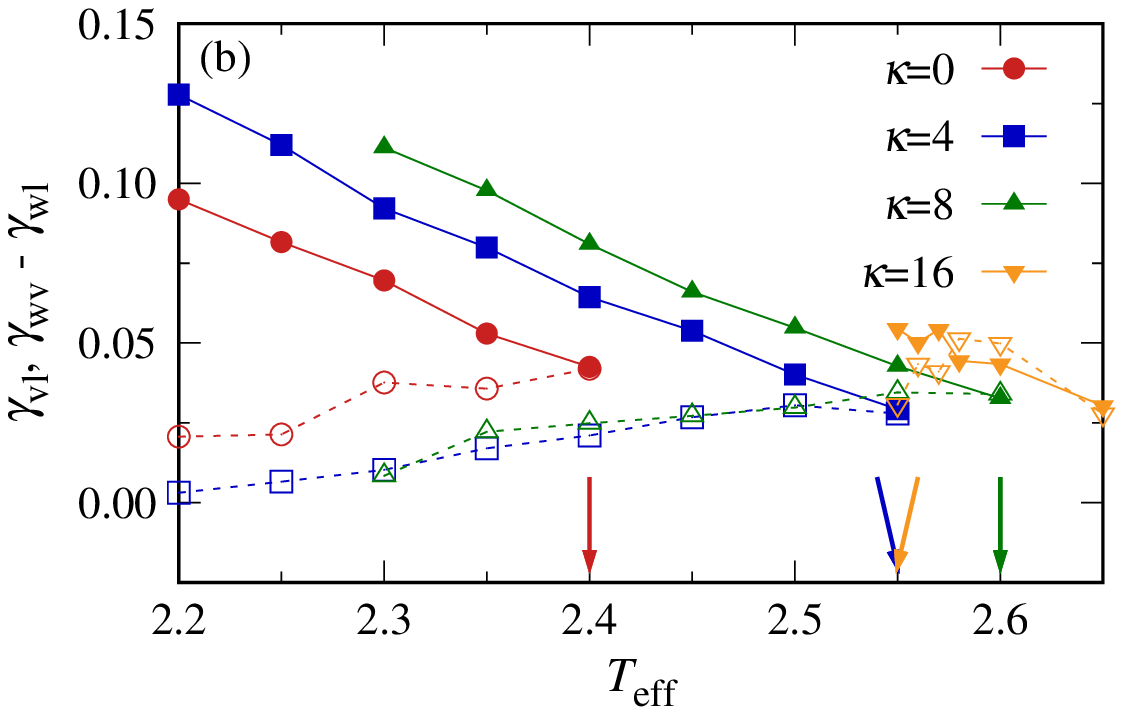}
    \caption{(a) $\gamma_{\rm vl}$ (filled symbols) and $\gamma_{\rm wv}-\gamma_{\rm wl}$ (open symbols) as functions of $T_{\rm eff}$ for flexible polymers at three different chain lengths $N=8$, $16$ and $32$. (b) Same plot as (a), but for different values of $\kappa=0$, $4$, $8$ and $16$ at fixed chain length $N=16$. The colored arrows indicate the location of the wetting transition temperature, $T_{\rm w}$.}
    \label{Fig:gamma_VL_gamma_diff}
\end{figure}

With the computed values of $\gamma_{\rm vl}$, $\gamma_{\rm wv}$ and $\gamma_{\rm wl}$ at different $T_{\rm eff}$, we can employ Young's equation [Eq. (\ref{Eq:Young})] to accurately determine the temperature of the wetting transition, $T_{\rm w}$. Figure~\ref{Fig:gamma_VL_gamma_diff}(a) shows $\gamma_{\rm vl}$ as well as $\gamma_{\rm wv}-\gamma_{\rm wl}$ as functions of $T_{\rm eff}$ for flexible polymers ($\kappa=0$) with different chain lengths $N=8$, $16$ and $32$. The cosine of the contact angle satisfies $\cos(\theta)<1$ in the partial wetting regime, so $\gamma_{\rm wv}-\gamma_{\rm wl} < \gamma_{\rm vl}$ there. The value of $\gamma_{\rm wv}-\gamma_{\rm wl}$ increases with increasing $T_{\rm eff}$ and finally coincides with $\gamma_{\rm vl}$ at complete wetting, so that $\cos(\theta)=1$. We have observed that $T_{\rm w}$ increases with increasing $N$, as indicated by the colored arrows. A qualitatively similar behavior is encountered also for a different model studied by DFT (see Appendix). For solutions of semiflexible polymers, the variation of $T_{\rm w}$ as functions of $\kappa$ is presented in Fig.~\ref{Fig:gamma_VL_gamma_diff}(b) for $N=16$. The overall behavior of $\gamma_{\rm vl}$ and $\gamma_{\rm wv}-\gamma_{\rm wl}$ vs $T_{\rm eff}$ is similar to the case of flexible polymers. The wetting temperature $T_{\rm w}$ increases with increasing $\kappa$, as the effective length of the polymers increases, which is similar to the behavior of the critical temperature $T_{\rm c}$. Note that for $\kappa = 16$ and $T_{\rm eff} < 2.55$ the stable liquid-like phase is nematically ordered. Due to the anisotropy of this phase we could not use the Kirkwood-Buff relation [Eq.~(\ref{Eq:KirkwoodBuff})] to obtain interfacial tensions for $T_{\rm eff} < T_{\rm in}$.

\begin{figure}[htb!]
    \centering
    \includegraphics[width=\figwidth]{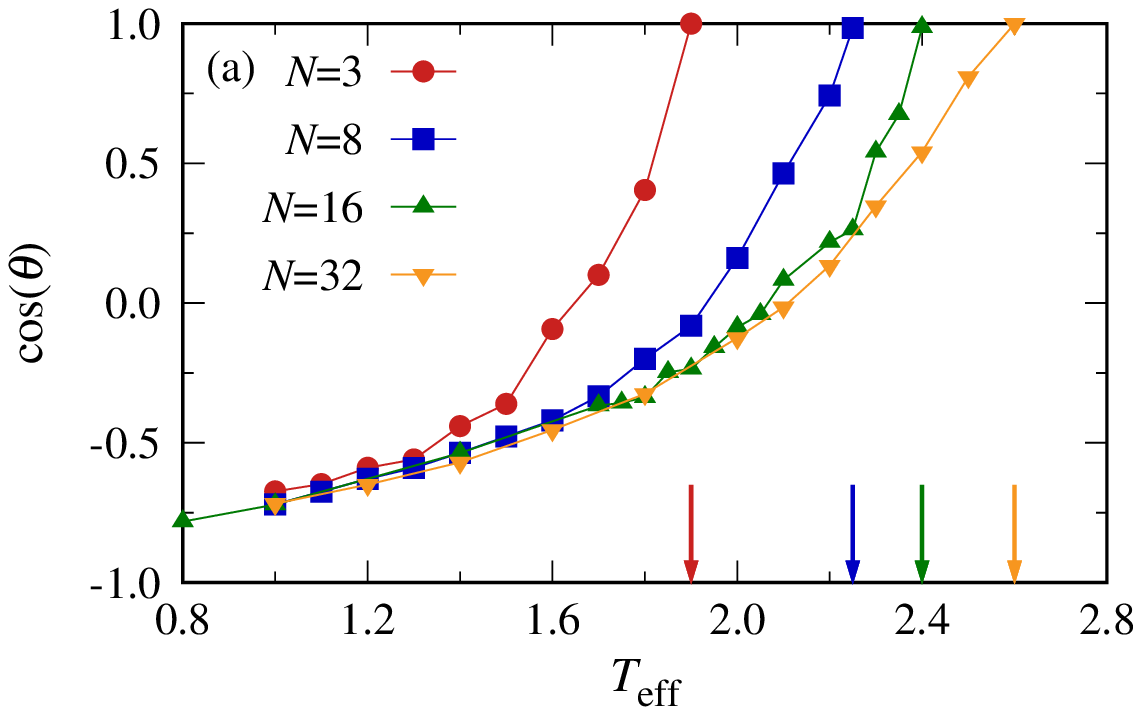}
    \includegraphics[width=\figwidth]{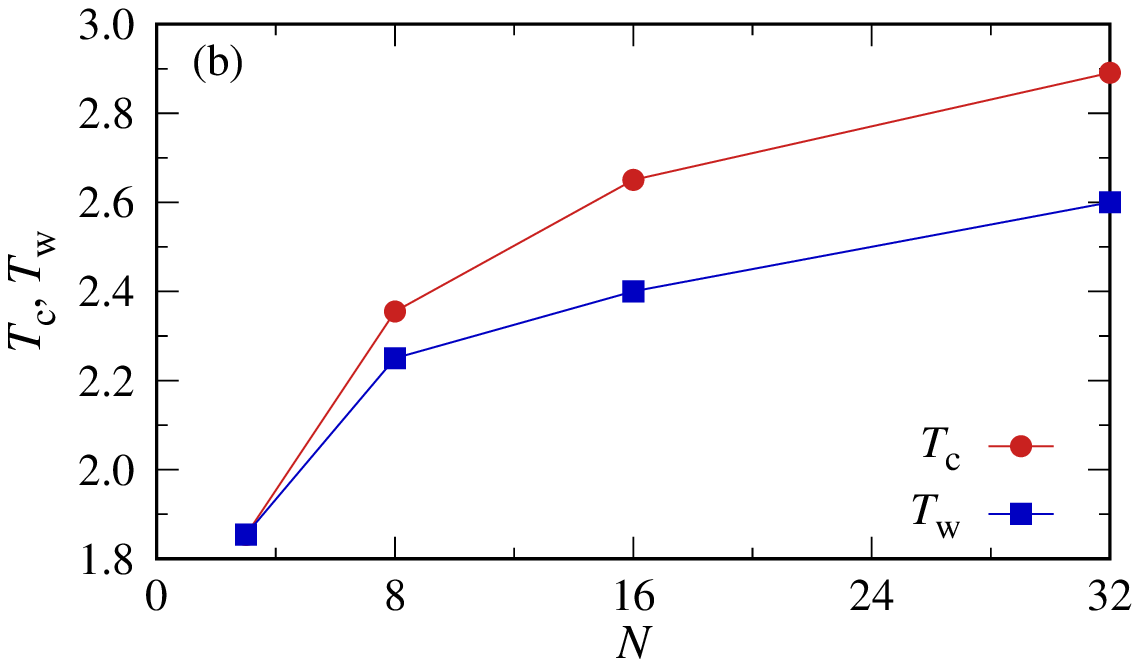}
    \caption{(a) Cosine of the contact angle $\cos(\theta)$ [Eq.~(\ref{Eq:Young})] vs $T_{\rm eff}$ for flexible polymers with $N=3$, $8$, $16$ and $32$. $\cos(\theta) = 1$ and $-1$ identify the locations of complete wetting and drying transitions, respectively. The colored arrows indicate the wetting transition temperatures, $T_{\rm w}$. (b) Variation of $T_{\rm c}$ and $T_{\rm w}$ as functions of $N$.}   
    \label{Fig:cosine+MD+DFT}
\end{figure}

For flexible polymers, $\cos(\theta)$ as functions of $T_{\rm eff}$ are presented in Fig. \ref{Fig:cosine+MD+DFT}(a). The value of $\cos(\theta)$ increases monotonically with increasing $T_{\rm eff}$ and finally reaches $\cos(\theta)=1$ at complete wetting. For trimers ($N=3$), complete wetting occurs exactly at the critical temperature, i.e., $T_{\rm w}=T_{\rm c}$. Figure~\ref{Fig:cosine+MD+DFT}(a) suggests that the wetting transition is of first order, although the behavior of $\Delta \rho_{\rm v}$ indicates that it is rather a weak first order transition. Further we have noticed that the value of $\cos(\theta)$ is negative for small $T_{\rm eff}$, which indicates partial drying. We were, however, unable to achieve complete drying [$\cos(\theta)=-1$] for the selected parameters, as $\cos(\theta) > -1$ at all temperatures where the polymer-rich liquid can still be observed in thermal equilibrium. In this low temperature regime, all curves from different $N$ collapsed onto a single master curve, which suggests that the contact angle $\theta$ is independent of $N$. In order to verify that this property is not a coincidence for the specific model studied here, we have investigated a different model by DFT (see Appendix), and found that also there the contact angle becomes independent of $N$ at low temperatures. The variation of $T_{\rm w}$ as a function of $N$ is shown in Fig. \ref{Fig:cosine+MD+DFT}(b) which increases monotonically with $N$, which is analogous to the behavior of $T_{\rm c}$. 

\begin{figure}[htbp!]
    \centering
    \includegraphics[width=\figwidth]{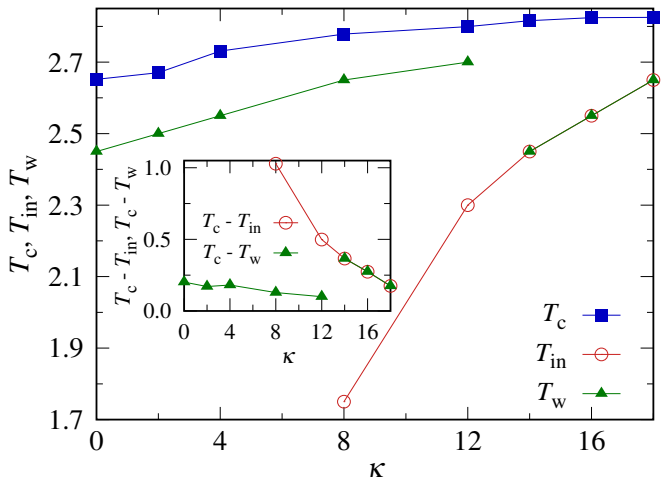}
    \caption{Critical temperature $T_{\rm c}$, isotropic-nematic transition temperature $T_{\rm in}$, and wetting transition temperature $T_{\rm w}$ vs bending stiffness $\kappa$. Inset: $T_{\rm c}-T_{\rm in}$ and $T_{\rm c}-T_{\rm w}$ vs $\kappa$ at $N=16$.}
    \label{Fig:Tw+Tc_vs_k+semiflex}
\end{figure}

For semiflexible polymers ($\kappa > 0$) with $N=16$, the variation of $T_{\rm w}$ as function of $\kappa$ is presented in Fig. \ref{Fig:Tw+Tc_vs_k+semiflex}. For $\kappa \leq 12$, the wetting temperature $T_{\rm w}$ increases monotonically with increasing bending stiffness (analogous to the behavior of $T_{\rm c}$), until it sharply drops near $\kappa* \approx 14$. Interestingly, for $\kappa \geq \kappa^*$, the wetting transition occurs exactly at the isotropic-nematic (i-n) transition, i.e. $T_{\rm w}=T_{\rm in}$. Then, both $T_{\rm w}$ and $T_{\rm in}$ increase with increasing $\kappa$. The inset of Fig.~\ref{Fig:Tw+Tc_vs_k+semiflex} shows $T_{\rm c}-T_{\rm in}$ and $T_{\rm c}-T_{\rm w}$ as functions of $\kappa$, demonstrating that $T_{\rm c}-T_{\rm in}$ decreases monotonically with increasing $\kappa$, since $T_{\rm in}$ increases faster than $T_{\rm c}$. In contrast, the behavior of $T_{\rm c}-T_{\rm w}$ is non-monotonic as a function of $\kappa$: For $\kappa < \kappa^*$, $T_{\rm c} - T_{\rm w}$ decreases slightly with increasing $\kappa$, then jumps up at $\kappa = \kappa^*$ to follow the progression of $T_{\rm c}-T_{\rm in}$ for $\kappa \geq \kappa^*$.

\begin{figure}[htbp!]
    \centering
    \includegraphics[width=\figwidth]{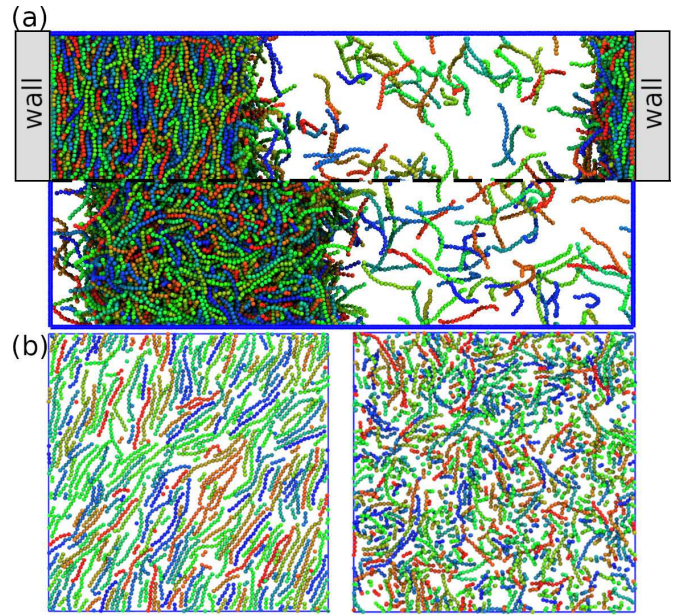}
    \caption{(a) Snapshot of semiflexible polymers with $\kappa = 16$ and $N=16$ at $T_{\rm eff}=2.55$ in (top) confined and (bottom) slab systems. (b) Cross-section view (in the $xy$ plane) inside the liquid region for (left) confined and (right) slab systems at $T_{\rm eff}=2.55$.}
    \label{Fig:comp-snaps+densProf+Teff2.55}
\end{figure}

\begin{figure}[htbp!]
    \centering
    \includegraphics[width=\figwidth]{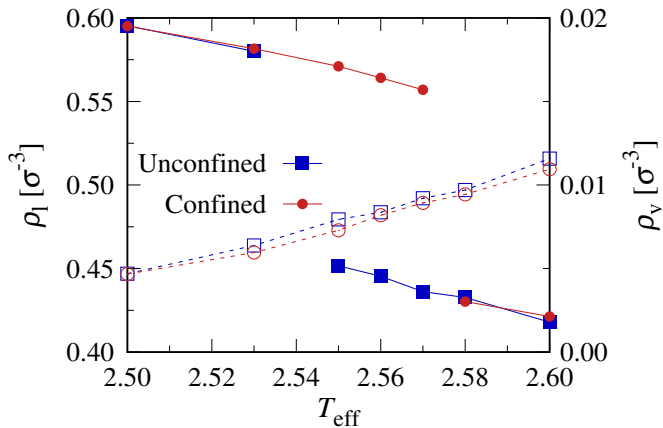}
    \caption{Coexisting densities in the liquid ($\rho_{\rm l}$, filled symbols and solid lines, left axis) and vapor ($\rho_{\rm v}$, open symbols and dashed lines, right axis) regions as functions of $T_{\rm eff}$ for semiflexible chains with $N=16$ and $k=16$. Data for the confined and unconfined systems shown in red and blue, respectively.}
    \label{Fig:comp_rhoL_vs_Teff_N16}
\end{figure}

The non-monotonic behavior of $T_{\rm w}$ for varying $\kappa$ can be better understood by considering the system with $N=16$ and $\kappa=16$ in more detail. Figure~\ref{Fig:comp-snaps+densProf+Teff2.55}(a) shows a snapshot for the confined system at $T_{\rm eff}=2.55$, which roughly corresponds to the i-n transition temperature for the corresponding unconfined system (Fig.~\ref{Fig:phase_diagram}).\cite{JMidyaJChemPhys2019} At this state point, we find in the confined systems the coexistence of a low density vapor phase with a high density nematic liquid phase. Interestingly, the liquid phase in the confined systems has a much higher density compared to the dense phase from our slab simulations of vapor-liquid coexistence (Fig.~\ref{Fig:comp_rhoL_vs_Teff_N16}), indicating that the semiflexible chains are packed more tightly in confinement. One might suspect that the observed higher density corresponds to the nematic branch of the phase diagram. To check this conjecture, cross-sectional views of the liquid regions are shown in Fig. \ref{Fig:comp-snaps+densProf+Teff2.55}(b) for the confined as well as for the slab systems. In the confined case, nematic order of the chains is observed, whereas a random orientation of the chains is observed in the unconfined system, confirming the coexistence of the isotropic liquid with its vapor phase. This difference indicates that spatial confinement shifts the i-n transition temperature $T_{\rm in}$ to a slightly higher value, which is the effect of capillary nematization.\cite{nikoubashman:jcp:2021} Exactly at $T_{\rm in}$, there is a triple point with three phases coexisting in thermal equilibrium, namely bulk vapor, bulk isotropic liquid, and bulk nematic phase. Thus, it is plausible that a wall then favors vapor-nematic coexistence.

We determined the coexistence densities near $T_{\rm in}$ and compared the results with the corresponding slab simulations (Fig.~\ref{Fig:comp_rhoL_vs_Teff_N16}). For the confined systems, there exists a small temperature window over which we found the coexistence of a nematic liquid with an isotropic vapor phase, in contrast to the unconfined systems which exhibit isotropic vapor-liquid coexistence over the investigated temperature range. This finding corroborates that the presence of walls shifts $T_{\rm in}$ to higher values due to capillary nematization. In contrast, the density of the vapor phase is barely affected due to confinement.

\begin{figure}[htbp!]
    \centering
    \includegraphics[width=\figwidth]{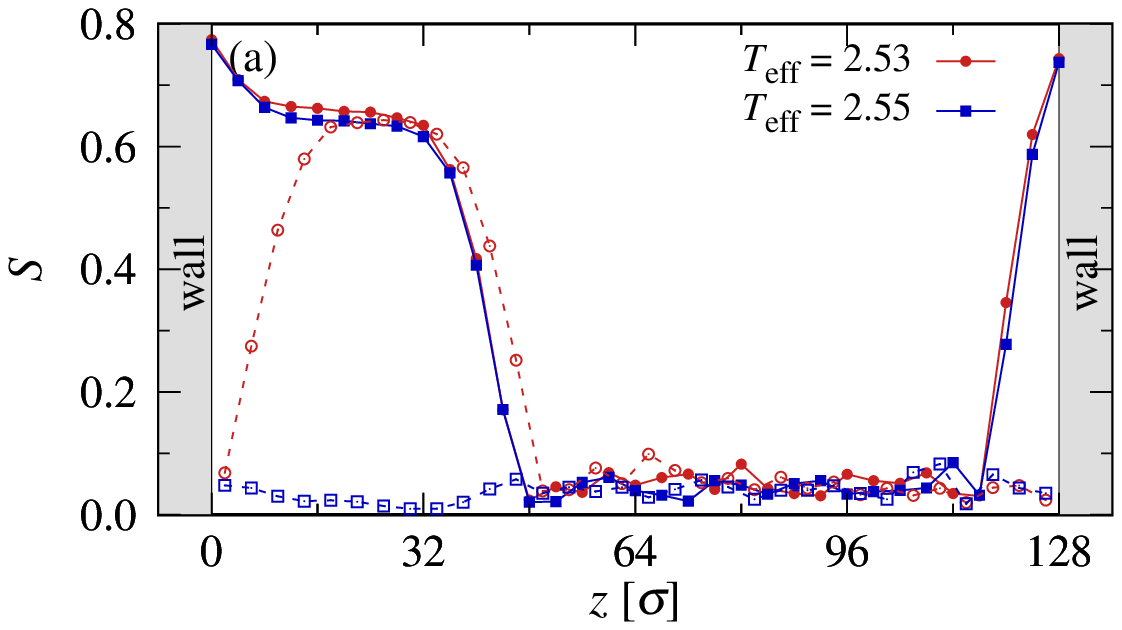}
    \includegraphics[width=\figwidth]{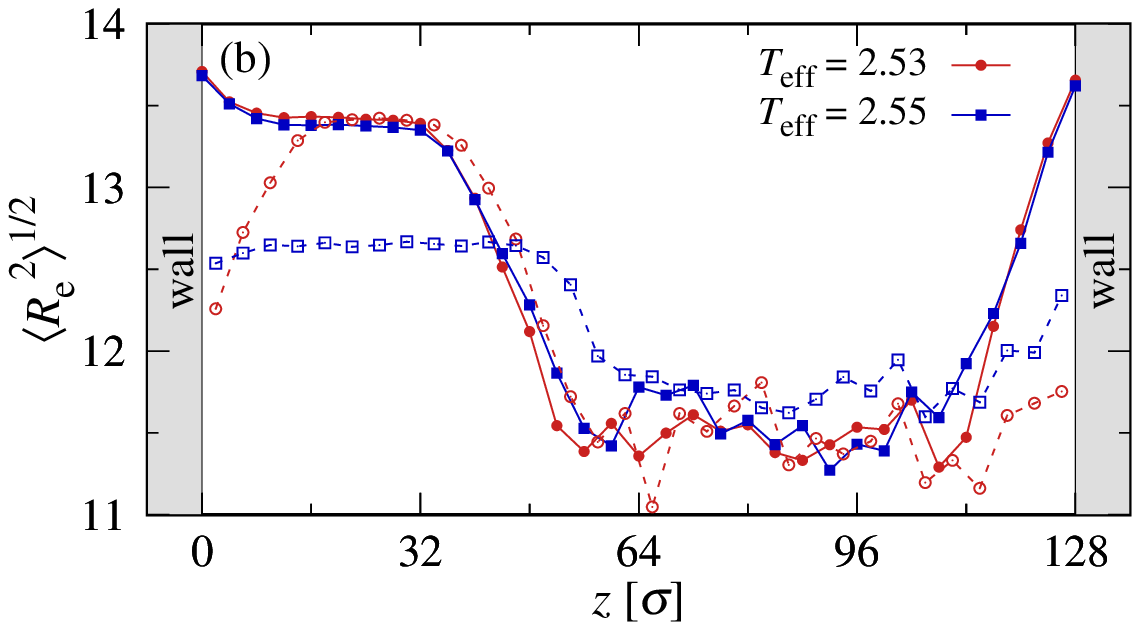}
    \includegraphics[width=\figwidth]{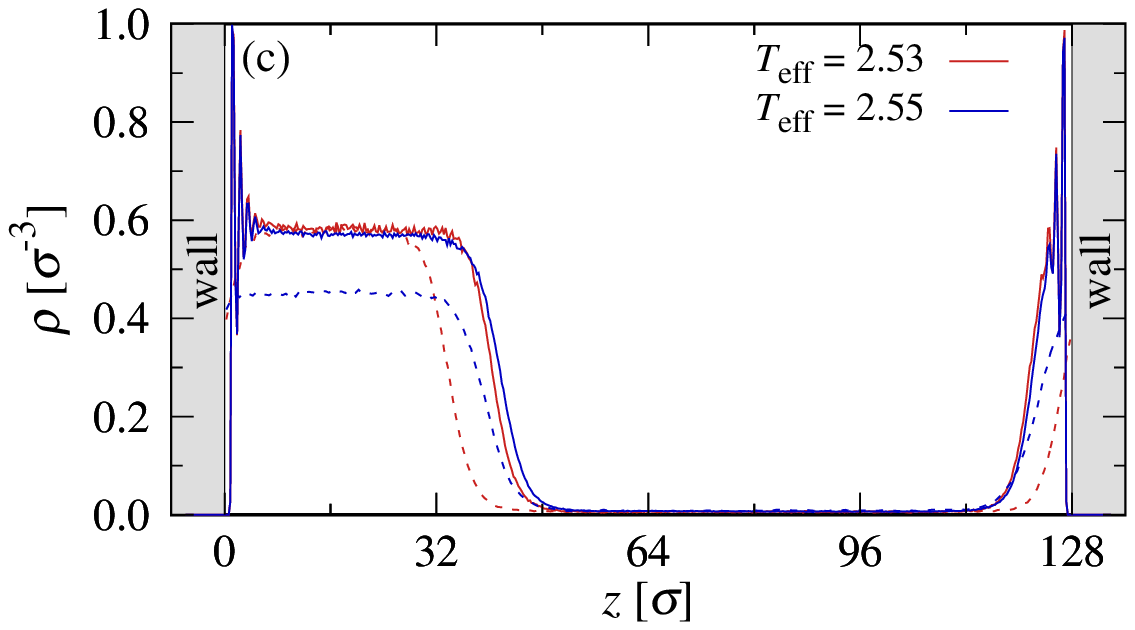}
    \caption{(a) Local nematic order parameter $S$ as functions of center of mass position $z$ for confined (solid lines and filled symbols) and unconfined (dashed lines and open symbols) systems with $N=16$ and $\kappa=16$ at different $T_{\rm eff}$, as indicated. The small positive values of $S$ in isotropic regions are due to binning. (b) Root mean square end-to-end distance $\sqrt{\langle R_{\rm e}^2 \rangle}$ as functions of $z$. (c) Monomer number density $\rho$ as functions of $z$. Making use of the translational invariance of systems with periodic boundary conditions, the unconfined systems have been placed such that the flat regions of the liquid film density profiles fall also in the region $0 < z/\sigma < 40$, where they occur for the confined systems.}
    \label{Fig:OrienOP+e2eDist+deltaTeff}
\end{figure}

To confirm the (local) nematic order of the chains, we compute the orientational order-parameter $S$, which is the largest eigenvalue of the tensor
\begin{equation}
    \mathcal{Q}_{i,n}^{\alpha \beta} = \frac{1}{2} (3 {\bf u}_{i,n}^{\alpha} {\bf u}_{i,n}^{\beta}-\delta_{\alpha \beta}),
    \label{Eq:orien_OP}
\end{equation}
with ${\bf u}_{i,n}$ being the unit vector connecting monomers $i$ and $i+1$ of the $n^{\rm th}$ chain. The local nematic order in a slab of thickness $\Delta z$ is determined by considering all bond vectors ${\bf u}_{i,n}$ of chains whose center of mass lies in between $z$ and $z+\Delta z$. Figure \ref{Fig:OrienOP+e2eDist+deltaTeff}(a) shows $S$ as a function of chain position $z$. For $T_{\rm eff}=2.53$, a temperature just below the i-n transition for bulk systems, the large value of $S$ indicates that the chains are ordered nematically inside the liquid slab, for both systems. For $T_{\rm eff}=2.55$, we find $S \simeq 0$ and thus an isotropic liquid in the unconfined system, whereas the confined system has $S \approx 0.64$ in the dense phase, confirming nematic order of the chains. 

\begin{figure}[htbp!]
    \centering
    \includegraphics[width=\figwidth]{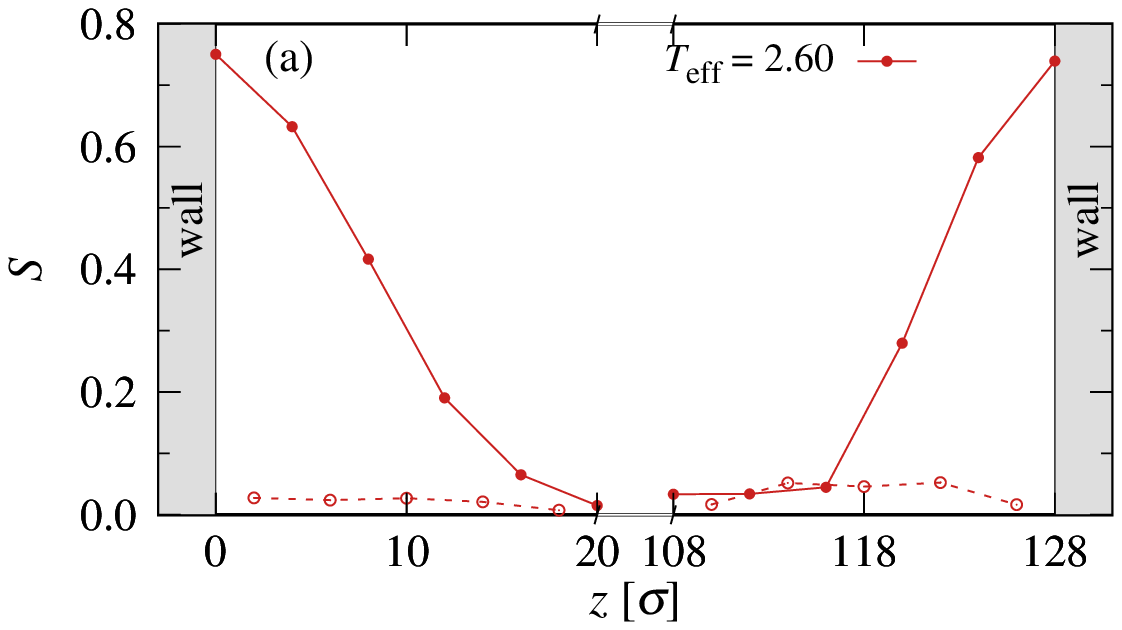}
    \includegraphics[width=\figwidth]{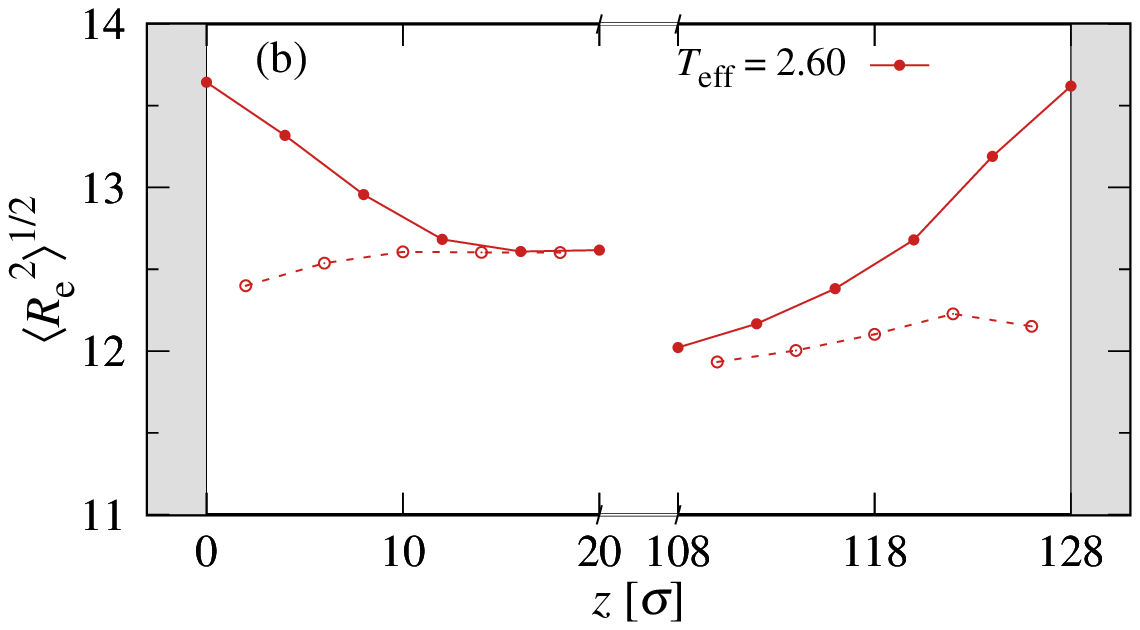}
    \includegraphics[width=\figwidth]{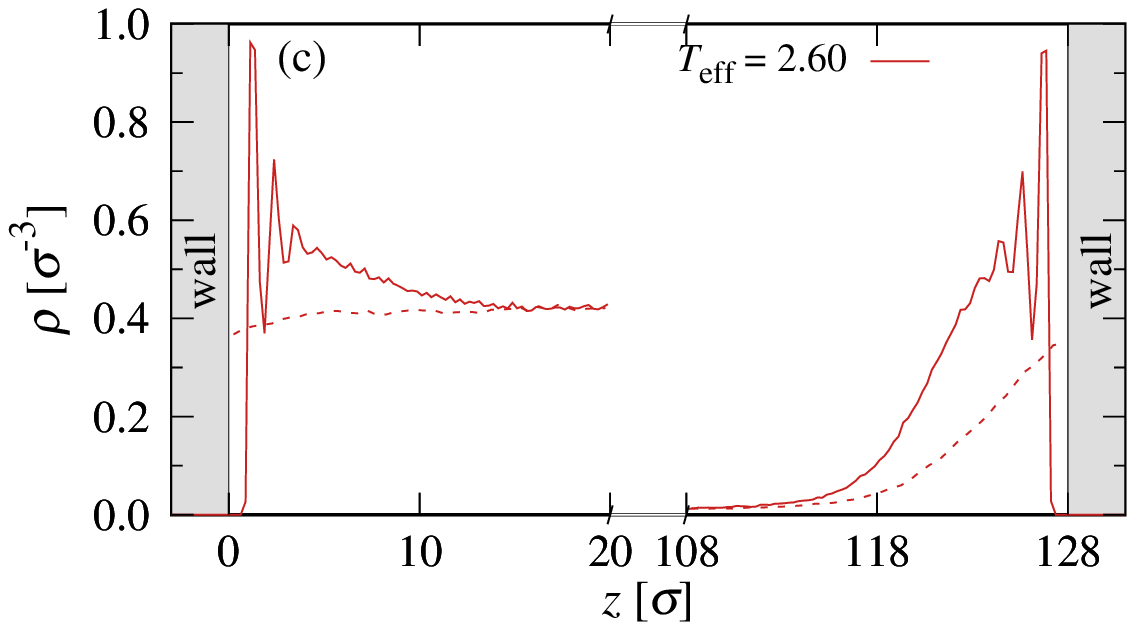}
    \caption{Same as Fig.~\ref{Fig:OrienOP+e2eDist+deltaTeff}, but for $T_{\rm eff}=2.60$. Note the larger scale for $z$ and the broken abscissa range.}
    \label{Fig:OrienOP+e2eDist+deltaTeff+T2.60}
\end{figure}

With the further increase $T_{\rm eff}$, we observe coexistence of an isotropic liquid with vapor for both systems, which is demonstrated for $T_{\rm eff} = 2.60$ in Fig.~\ref{Fig:OrienOP+e2eDist+deltaTeff+T2.60}. However, a very thin layer of nematically ordered polymers is still found close to walls due to capillary nematization. The thickness of this nematic layer is about $d_{\rm n} \approx 9\,\sigma$ for the liquid side [Fig.~\ref{Fig:OrienOP+e2eDist+deltaTeff+T2.60}(a), left part] and $d_{\rm n} \approx 7\,\sigma$ for the vapor side [Fig.~\ref{Fig:OrienOP+e2eDist+deltaTeff+T2.60}(a), right part]. Similar behavior is observed when we plot the root mean square end-to-end distance, $\sqrt{\langle R_{\rm e}^2 \rangle}$ of the chains as function of their center of mass position $z$, see Fig. \ref{Fig:OrienOP+e2eDist+deltaTeff+T2.60}(b). On about the same length scales, the monomer number density $\rho$ [Fig.~\ref{Fig:OrienOP+e2eDist+deltaTeff+T2.60}(c)], which exhibits strong layering with 3 clearly resolved peaks at both sides, decays to about $\rho_{\rm l} = 0.42\,\sigma^{-3}$ in the liquid but to a much smaller value in the vapor phase ($\rho_{\rm v} = 0.012\,\sigma^{-3}$, see Fig.~\ref{Fig:comp_rhoL_vs_Teff_N16}). This strong layering in the vapor phase with a large value of $S$ in the first layer adjacent to the wall can be expected to persist to significantly larger values of $T_{\rm eff}$: With the chosen strong attraction to the wall, even for good solvent conditions a quasi-two-dimensional almost nematic layer is found for similar choices of the parameters $\kappa$ and $N$.\cite{AMilchevSoftMatter2017} We speculate that also in the present case this nematic ``order'' is no true long range order, but one should rather find a power-law decay of nematic correlations up to $T_{\rm eff, KT}$, where a Kosterlitz-Thouless (KT) transition of this layer to an isotropic phase occurs. No such behavior is expected for much smaller choices of $\kappa$, where the nematic phase exists only for $T_{\rm eff}$ much smaller than $T_{\rm w}$.

\section{Conclusions}
\label{sec:conclusions}
In this work, the wetting behavior of polymer solutions is studied by molecular dynamics (MD) simulations of a coarse-grained model where the quality of the (implicit) solvent is varied. We treat fully flexible macromolecules, where the number of monomers is varied from oligomers ($N = 3, 8$) up to short polymers ($N=16, 32$), as well as semiflexible polymers with $N=16$ beads at varying stiffness $\kappa$. Wetting in a solution of flexible polymers is also studied by density functional theory (DFT) for a similar model of short polymer chains, but using a longer-ranged monomer-wall attraction. Qualitatively similar results are obtained by both methods, suggesting that the observed wetting behavior is qualitatively insensitive to the details of the studied models.

A distinctive methodological aspect of our work is the use of large enough linear dimensions in the MD simulations, so that a system with three separate, non-interacting interfaces can be studied (wall-liquid, liquid-vapor, and vapor-wall, Fig.~\ref{Fig:snap+densProf_WVL}). The use of the Kirkwood-Buff relation for the pressure tensor anisotropy then yields the sum of $\gamma_{\rm wl} + \gamma_{\rm vl} + \gamma_{\rm wv}$. The surface tensions $\gamma_{\rm vw}$ and $\gamma_{\rm vl}$ can be separately found in a similar fashion (see Fig.~\ref{Fig:snap+densProf+WV}), and so the contact angle $\theta$ can be directly inferred using Young's equation, with no need to actually study droplets.

We find for the MD and DFT models, that the wetting transition is of first order (Fig.~\ref{Fig:gamma_VL_gamma_diff}), and that the wetting transition temperature $T_{\rm w}$ increases with increasing $N$, but somewhat slower than the critical temperature $T_{\rm c}$ of the bulk liquid-vapor type transition (Fig.~\ref{Fig:cosine+MD+DFT}). For $T \ll T_{\rm w}$ we find partial drying ($\theta > 90^\circ$). The MD results settle down at low temperatures at $N$-independent values of the contact angle [Fig.~\ref{Fig:cosine+MD+DFT}(a)]. Also the DFT calculations predict a similar increase of the contact angle when the temperature is far below the wetting transition. In the MD work for $N = 16$ or larger, the wetting transition was found to be only weakly of first order: Approaching this transition from the partial wetting side, the adsorbed amount of fluid (i.e., the surface excess observed on the wall from the saturated vapor) is found to increase by about an order of magnitude when $T$ approaches $T_{\rm w}$ from below (Fig.~\ref{Fig:excess}). Note that theory predicts for long-range surface forces a wetting transition of first order always, while critical wetting could occur for a short-range wall potential. Since our wall potential is only cut off at a distance of five monomer diameters, where the potential is very small already, we can consider our wall potential as being long-ranged for all practical purposes.

Solutions of semiflexible polymers also exhibit a transition from an isotropic liquid to a nematic liquid in addition to the vapor-liquid transition. We have determined the phase diagram in the bulk in the case where the contour length and the persistence length of the polymers are almost equal (Fig.~\ref{Fig:phase_diagram}). The surface tension between the vapor and the isotropic liquid $\gamma_{\rm vl}$ shows an Ising-like critical behavior near the vapor-liquid critical temperature, irrespective of chain stiffness (Fig.~\ref{Fig:norm_gamma_VL}). The behavior of the difference of the wall tensions between vapor and isotropic liquid as function of temperature for different stiffnesses is similar to the case of flexible polymers (Fig.~\ref{Fig:gamma_VL_gamma_diff}). The wetting transition temperature increases with increasing chain stiffness similar as $T_{\rm c}$ does up to $\kappa = 12$, while for $\kappa \geq 14$ wetting occurs at the (considerably smaller) triple point temperature where the nematic phase in the bulk can coexist with both vapor and isotropic liquid (Fig.~\ref{Fig:Tw+Tc_vs_k+semiflex}). Note that nematic phases occur for the present model under good solvent conditions only for $\kappa \geq 8$. The presumably singular variation of $T_{\rm w}$ with $\kappa$ (Fig.~\ref{Fig:Tw+Tc_vs_k+semiflex}) is a very interesting finding, and only speculations on the origin of this behavior can be offered; when the temperature is raised for $\kappa = 16$, the vapor-liquid equilibrium changes its character discontinuously at the isotropic-nematic transition temperature $T_{\rm in}$ from vapor-nematic liquid to vapor-isotropic liquid (Fig.~\ref{Fig:phase_diagram}). This discontinuous change may give rise to discontinuities in the vapor-liquid interfacial tensions and/or the difference of wall tensions. Consequently, one expects discontinuous jumps in the contact angle if $T_{\rm in}$ is much smaller than $T_{\rm w}$, while close to $T_{\rm w}$ one can expect that the contact angle becomes zero at $T_{\rm w}$. While at $T_{\rm eff}$ somewhat higher than $T_{\rm in}$ the wall can stabilize a nematic layer of microscopic thickness only, this nematically ordered region is expected to become macroscopic as $T_{\rm eff}$ decreases toward $T_{\rm in}$.

\section*{Acknowledgments}
This work was funded by the Deutsche Forschungsgemeinschaft (DFG, German Research Foundation) through projects 261177998; 274340645; NI 1487/7-1. S.A.E. thanks the Alexander von Humboldt Foundation for support. The authors gratefully acknowledge the computing time granted on the supercomputer Mogon (hpc.uni-mainz.de).

\appendix
\section{Density Functional Theory}
A tangent-sphere model is used for the polymers in our DFT calculations, with bond length $\ell_{\rm b} = \sigma$ and non-bonded interactions given by the LJ potential [Eq.~(\ref{eq:ULJ})]. The interaction between the monomers and the flat walls is modeled via\cite{yatsyshin18}
\begin{equation}
    U_{\rm wall}(z)=4\pi\rho_{\rm w}\varepsilon_{\rm w}\sigma_{\rm w}^{3}
    \left[\frac{1}{45}\left(\frac{\sigma_{\rm w}}{H_0+z}\right)^{9}
    -\frac{1}{6}\left(\frac{\sigma_{\rm w}}{H_0+z}\right)^{3}\right],  
\label{phiwp}
\end{equation}
where $z$ is the distance of a monomer from the wall, $\rho_{\rm w}=\sigma^{-3}$ is the particle density of the substrate, $H_0=5$ is the wall coating parameter which is introduced to avoid numerical problems associated with the divergence of $U_{\rm wall}(z)$ at $z=0$, $\varepsilon_{\rm w}=0.35\,\varepsilon$ sets the interaction strength, and $\sigma_{\rm w}=2\,\sigma$ is the diameter of the (fictitious) wall beads.\cite{yatsyshin18}

As a starting point of the DFT-based treatment, the grand free energy $\Omega$ is expressed as a functional of the polymer density profile $\rho_{\rm p}(\mathbf{R}_{\rm p})$, where $\mathbf{R}_{\rm p}=(\mathbf{r}_1, \mathbf{r}_2,\cdots,\mathbf{r}_N)$ is a collective variable with the individual monomer coordinates $\mathbf{r}_i$. The minimization of $\Omega$ with respect to $\rho_{\rm p}(\mathbf{R}_{\rm p})$ yields the equilibrium polymer density distribution.  The functional $\Omega$ is related to the Helmholtz free energy functional $F$ through a Legendre transform
\begin{equation}
    \Omega[\rho_{\rm p}(\mathbf{R}_{\rm p})]=F[\rho_{\rm p}(\mathbf{R}_{\rm p})]+\int d\mathbf{R}_{\rm p}
    \rho_{\rm p}(\mathbf{R}_{\rm p}))[V_{\rm ext}(\mathbf{R}_{\rm p})-\mu],
\label{omega}
\end{equation}
where $\mu$ is the polymer chemical potential, and $V_{\rm ext}(\mathbf{R}_{\rm p})$ is the external field due to the interaction of the polymer beads with the wall 
\begin{equation}
    V_{\rm ext}(\mathbf{R}_{\rm p})=\sum_{i=1}^{N} U_{\rm wall}(\mathbf{r}_i).
    \label{vext}
\end{equation}

We employ the following approximation for $F$, which separates it into ideal and excess parts\cite{SAEgorovPRE2004} 
\begin{equation}
    F[\rho_{\rm p}(\mathbf{R}_{\rm p})] = F_{\rm id}[\rho_{\rm p}(\mathbf{R}_{\rm p})] +F_{\rm ex}[\rho(\mathbf{r})],  
    \label{ftot}
\end{equation}
with the ideal functional given by~\cite{IChubakJPCB2021, CEWoodwardJCP1991} 
\begin{equation}
    \beta F_{\rm id}[\rho_{\rm p}(\mathbf{R}_{\rm p})] =
    \int {\rm d}\mathbf{R}_{\rm p} \rho_{\rm p}(\mathbf{R}_{\rm p}))[\ln \rho_{\rm p}(\mathbf{R}_{\rm p})-1]
    +\beta \int {\rm d}\mathbf{R}_{\rm p} \rho_{\rm p}(\mathbf{R}_{\rm p}) V_{\rm b}(\mathbf{R}_{\rm p})
\label{fideal}
\end{equation}
with $\beta=1/(k_{\rm B}T)$, and binding energy $V_{\rm b}(\mathbf{R}_{\rm p})$ given by\cite{SAEgorovPRE2005}
\begin{equation}
\exp[-\beta V_{\rm b}(\mathbf{R}_{\rm p})]=\prod_{i=1}^{N-1}\frac
    {\delta(|\mathbf{r}_i-\mathbf{r}_{i+1}|-\sigma)}{4\pi\sigma^{2}}
    =\prod_{i=1}^{N-1}g_{\rm b}(|\mathbf{r}_i-\mathbf{r}_{i+1}|). 
\label{vbond}  
\end{equation}

The excess term is written as a functional of the monomer density given by\cite{IChubakJPCB2021, CEWoodwardJCP1991}
\begin{equation}
    \rho(\mathbf{r})=\int {\rm d}\mathbf{R}_{\rm p}\sum_{i=1}^{N}
    \delta(\mathbf{r}-\mathbf{r}_i)\rho_{\rm p}(\mathbf{R}_{\rm p}),
\label{rhor}
\end{equation}
which is split into a repulsive and an attractive term\cite{SAEgorovPRE2004}
\begin{equation}
    F_{\rm ex}[\rho(\mathbf{r})]=F_{\rm rep}[\rho(\mathbf{r})]+F_{\rm att}[\rho(\mathbf{r})]. 
\label{fex}
\end{equation}
For the former, we adopt the weighted density approximation\cite{SAEgorovJCP2008} 
\begin{equation}
    \beta F_{\rm rep}[\rho(\mathbf{r})]=\int {\rm d}\mathbf{r} \rho(\mathbf{r}) f_{\rm rep}(\bar{\rho}(\mathbf{r})),
\label{fexcess}   
\end{equation}
with weighted density
\begin{equation}
    \bar{\rho}(\mathbf{r})=\int {\rm d}{\mathbf{r}}^{\prime}
    \rho(\mathbf{r}^{\prime})w(|\mathbf{r}-\mathbf{r}^{\prime}|),
\label{rhobar}   
\end{equation}
where $f_{\rm rep}(\rho)$ is the excess free energy density per site of the polymer melt with site density $\rho$, arising from the short-ranged hard-core repulsive interactions. We compute it from  Wertheim's expression which was obtained on the basis of first-order thermodynamic perturbation theory\cite{MSWertheimJCP1987}
\begin{equation}
    f_{\rm rep}(\rho)=\frac{4\eta-3\eta^2}{(1-\eta)^2}-\left(1-\frac{1}{N}\right)\ln{\frac{1-\eta/2}{(1-\eta)^3}}
\label{fextpt1}
\end{equation}
with monomer packing fraction $\eta=\pi \sigma^{3}\rho/6$.

We employ the simple square-well form for the weighting function $w(r)$,  
whose range is given by the diameter $\sigma$ of the polymer
segment\cite{FLoVersoMacromolecules2012}
\begin{equation}
    w(r)=\frac{3}{4\pi\sigma^{3}}\Theta(\sigma-r),
\label{weight}
\end{equation} 
with Heaviside step function $\Theta(r)$. While more sophisticated forms of the weighting function are available in the literature
(e.g., those used in the Fundamental Measure Theory version of DFT~\cite{RRothJPCM2010}), previous studies\cite{MTuressonPRE2007} have shown relative insensitivity of DFT results for polymeric systems to the specific choice of the weight function. 

Regarding the attractive contribution to the excess free energy,\cite{MMuellerJChemPhys2003} in our earlier DFT study of nanoparticle interactions in a polymer melt,\cite{NPatelJCP2005} we have found that accurate results for polymer density profiles were obtained using a simple mean-field approximation for the attractive part of $F_{\rm ex}$
\begin{equation}
    F_{\rm att}[\rho(\mathbf{r})]= \frac{1}{2} \int {\rm d}\mathbf{r}\int {\rm d}\mathbf{r}^{\prime}\rho(\mathbf{r}) \rho(\mathbf{r}^{\prime}) U_{\rm pp}^{\rm att}(|\mathbf{r}-\mathbf{r}^{\prime}|),
\label{fatt}
\end{equation}
with\cite{yatsyshin18}
\begin{equation}
    U_{\rm pp}^{\rm att}(r) = \begin{cases}
    0 & ,r\leq \sigma\\
    U_{\rm LJ}(r) & , \text{else} .
    \end{cases}
\label{vrabatt}
\end{equation}
The minimization of the grand free energy functional $\Omega$ yields the following result for the equilibrium polymer density profile\cite{SAEgorovJChemPhys2016}
\begin{equation}
  \rho_{\rm p}(\mathbf{R}_{\rm p})=\rho_{\rm b}\prod_{i=1}^{N-1}g_{\rm b}(|\mathbf{r}_i-\mathbf{r}_{i+1}|)\prod_{i=1}^{N}\exp[-\lambda(\mathbf{r}_i)], 
\label{rhoprp}
\end{equation}
where
\begin{equation}
    \lambda(\mathbf{r})=\beta \frac{\delta F_{\rm ex}}{\delta \rho(\mathbf{r})}+\beta U_{\rm wall}(\mathbf{r}),
\label{lambdarp}
\end{equation}
and $\rho_{\rm b}$ is the bulk monomer density.  Substitution of $\rho_{\rm p}(\mathbf{R}_{\rm p})$ into Eq.~(\ref{rhor}) then yields an integral equation for the monomer density distribution $\rho(z)$ which was solved numerically on an equidistant grid with grid spacing $\Delta z =0.02\,\sigma$. Simple Picard iteration procedure was employed,\cite{SAEgorovJCP1997} and the tolerance criterion for
terminating the iterative procedure was set to $10^{-6}$. 

\begin{figure}[htbp]
    \includegraphics[width=\figwidth]{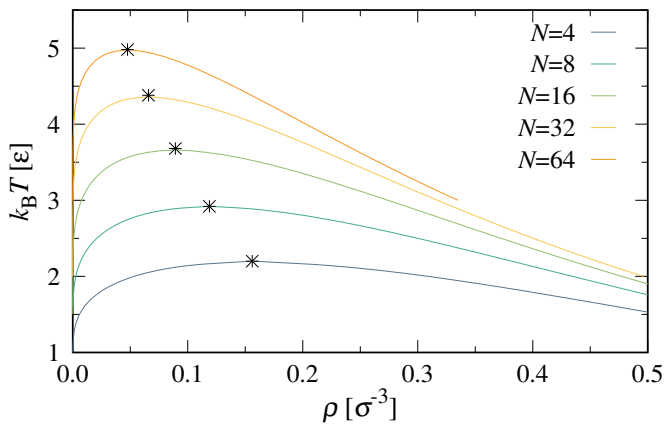}
    \caption{Vapor-liquid coexistence curves for fully flexible polymers of length $N$ in the $\rho-T$ plane. Critical points ($T_{\rm c}$, $\rho_{\rm c}$) are indicated by stars.}
\label{dft_fig1}
\end{figure}

We first determine the vapor-liquid coexistence curves for the flexible polymers (Fig.~\ref{dft_fig1}),\cite{JMidyaJChemPhys2019} by requiring the equality of pressure and chemical potential of the two coexisting phases, which yields the vapor and liquid monomer densities at coexistence, $\rho_{\rm v}$ and $\rho_{\rm l}$, respectively. Due to the mean-field nature of DFT,\cite{REvansAdvPhys1979, REvans:Book1992} its results for the near-critical region of the phase diagram are not reliable. In particular, the critical temperature $T_{\rm c}$ is overestimated, while the critical density $\rho_{\rm c}$ is underestimated compared to the MD simulation results (cf. Fig.~\ref{Fig:phase_diagram}).

Having determined $\rho_{\rm v}$ and $\rho_{\rm l}$, we now proceed to calculate the vapor-liquid surface tension $\gamma_{\rm vl}$. Here we assume that the vapor-liquid interface is perpendicular to the $z$-axis and the boundary conditions are set such that one has bulk vapor phase at small $z$ and bulk liquid phase at large $z$. From the resulting DFT density profiles and Eq.~(\ref{omega}), one can readily compute the grand potential density $\beta\omega(z)$, which yields $\gamma_{\rm vl}$
\begin{equation}
    \beta\sigma^2\gamma_{\rm vl}=\int_{-\infty}^{\infty}{\rm d}z\beta[\omega(z)-\omega_{\rm b}],
    \label{gammalv}
\end{equation}  
where $\beta\omega_{\rm b}$ is the bulk value of the grand potential density.

\begin{figure}[htbp]
    \includegraphics[width=\figwidth]{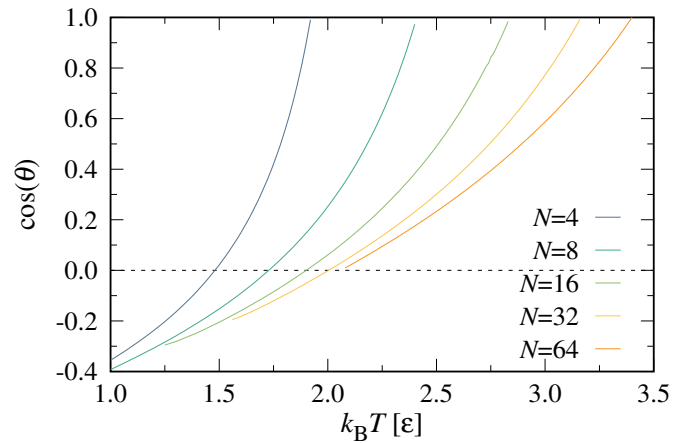}
    \caption{Cosine of the contact angle, $\cos(\theta)$, vs temperature $T$ from DFT for several chain lengths $N$, as indicated.}
    \label{dft_fig2}
\end{figure}

The wall-vapor and wall-liquid interfacial tensions are computed in a similar fashion,\cite{SAEgorovJCP2020} with an attractive flat wall [Eq.~(\ref{phiwp})] located in the $xy$-plane at $z=0$, and either vapor or liquid bulk density (at coexistence) imposed as the boundary condition for large $z$. From the equilibrium monomer profiles, one can again compute the grand potential density $\beta\omega(z)$, which yields the wall-vapor interfacial tension\cite{SAEgorovJCP2020}
\begin{equation}
    \beta\sigma^2\gamma_{\rm wv}=\int_{0}^{\infty}{\rm d}z\beta(\omega(z)-\omega_{\rm b}),
    \label{gammasl}
\end{equation}  
with the wall-liquid interfacial tension $\gamma_{\rm wl}$ obtained analogously. Having obtained $\gamma_{\rm vl}$, $\gamma_{\rm wv}$, and $\gamma_{\rm wl}$ for a range of temperatures, we can compute the contact angle $\theta$ from Young's equation [Eq.~(\ref{Eq:Young})]. The corresponding DFT results for $\cos(\theta)$ as functions of $T$ are shown in Fig.~\ref{dft_fig2} for several values of $N$. DFT is in qualitative but not in quantitative agreement with the simulations, due to different forms of the wall potential employed in the two methods, as well as inherent DFT approximations.

\begin{figure}[htbp]
    \includegraphics[width=\figwidth]{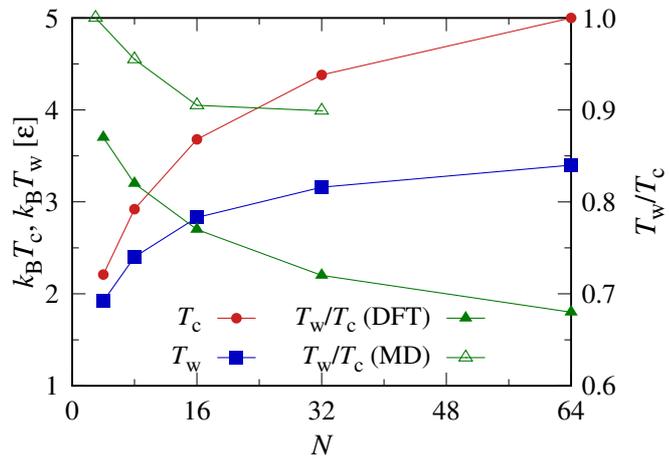}
    \caption{DFT results for the critical temperature $T_{\rm c}$, the wetting temperature $T_{\rm w}$ (left axis), and their ratio $T_{\rm w}/T_{\rm c}$ from DFT and MD (right axis), as functions of $N$.}
    \label{dft_fig3}
\end{figure}

Finally, Fig.~\ref{dft_fig3} provides a summary of the DFT results for the critical temperature $T_{\rm c}$, the wetting temperature $T_{\rm w}$ [where $\cos(\theta)=1$], and their ratio, as functions of $N$. Once again, there is qualitative agreement with the MD results.

\clearpage

\end{document}